\documentclass[a4paper,11pt]{article}
\usepackage{pos}

\usepackage{graphicx,epsfig,amsmath}
\usepackage{subcaption}
\usepackage{multirow}
\usepackage{slashed}
\usepackage{cleveref}
\usepackage{enumitem}
\usepackage{calc}
\usepackage{placeins}
\usepackage[compat=1.1.0]{tikz-feynman}
\usetikzlibrary{positioning}
\usepackage{xcolor}

\def\baligned{\begin{aligned}}
\def\ealigned{\end{aligned}}
    
\newcommand{\hc}{\text{H.c.}}

\def\be{\begin{equation}}
\def\ee{\end{equation}}
\def\bea{\begin{eqnarray}}
\def\eea{\end{eqnarray}}

\newcommand{\SM}{\text{SM}}
\newcommand{\SMEFT}{\text{SMEFT}}

\newcommand{\LSM}{{\cal L}_\SM}
\newcommand{\LSMEFT}{{\cal L}_\SMEFT}

\newcommand{\mhh}{m_{hh}}

\newcommand{\flambda}[1]{ \frac{#1}{\Lambda^2} }
\newcommand{\coeff}[2]{ \mathcal{C}_{#1} ^{#2} }
\newcommand{\CHbox}{\coeff{H\Box}{}}
\newcommand{\CHD}{\coeff{HD}{}}

\newcommand{\CHkin}{\coeff{H;\,\text{kin}}{ }}
\newcommand{\CH}{\coeff{H}{ }}
\newcommand{\CHG}{\coeff{HG}{ }}
\newcommand{\CtH}{\coeff{tH}{ }}
\newcommand{\CtG}{\coeff{tG}{ }}

\newcommand{\CQQ}{\coeff{QQ}{(1)}}

\newcommand{\CQQo}{\coeff{QQ}{(8)}}
\newcommand{\CQt}{\coeff{Qt}{(1)}}

\newcommand{\CQto}{\coeff{Qt}{(8)}}
\newcommand{\Ctt}{\coeff{tt}{ }}

\newcommand{\LHbox}{(\phi^{\dagger} \phi)\Box (\phi^{\dagger } \phi)}
\newcommand{\LHD}{(\phi^{\dagger} D_{\mu}\phi)^\ast(\phi^{\dagger}D^{\mu}\phi)}

\newcommand{\LH}{(\phi^{\dagger}\phi)^3}
\newcommand{\LHG}{(\phi^{\dagger} \phi)\, G_{\mu\nu}^a G^{\mu\nu,a}}
\newcommand{\LtH}{\left( (\phi^{\dagger}{\phi})(\bar{Q}_L t_R \tilde{\phi})+ \hc\right)}
\newcommand{\LtG}{\left( (\bar{Q}_L \sigma^{\mu\nu}T^a t_R \tilde{\phi})G^a_{\mu\nu}+ \hc\right)}
\newcommand{\LQQ}{(\bar{Q}_L\gamma^\mu Q_L)(\bar{Q}_L\gamma_\mu Q_L)}

\newcommand{\LQQo}{(\bar{Q}_L\gamma^\mu T^aQ_L)(\bar{Q}_L\gamma_\mu T^a Q_L)}
\newcommand{\LQt}{(\bar{Q}_L\gamma^\mu Q_L)\,\bar{t}_R\gamma_\mu t_R}
\newcommand{\LQto}{(\bar{Q}_L\gamma^\mu T^aQ_L)\bar{t}_R\gamma_\mu T^a t_R}
\newcommand{\Ltt}{\bar{t}_R\gamma^\mu t_R\,\bar{t}_R\gamma_\mu t_R}

\newcommand{\hcoeff}[1]{ c_{#1}}

\newcommand{\chhh}{\hcoeff{hhh}}

\newcommand{\order}[1] {\mathcal{O }\left( #1 \right)}

\pgfmathsetmacro\sizedot{1.1}
\pgfmathsetmacro\sizesqdot{2.0}
\pgfmathsetmacro\sizerectangle{0.8}
\pgfmathsetmacro\sizecirc{0.55}
\pgfmathsetmacro\sizecrodot{1.0}

\newcommand{\eqcomma}{\;,}

\newcommand{\MSbar}{\overline{\text{MS}}}

\title{Precise SMEFT predictions for di-Higgs production}

\author*[a]{Jannis Lang}

\affiliation[a]{Institute for Theoretical Physics, Karlsruhe Institute of Technology (KIT),\\
76131 Karlsruhe, Germany}

\emailAdd{jannis.lang@kit.edu}

\abstract{We present results of precision calculations for di-Higgs production 
that combine NLO QCD corrections with operators at canonical dimension six 
within Standard Model Effective Field Theory (SMEFT). 
We discuss possible options for operator contributions within a given EFT framework and sources of theory uncertainties.}

\FullConference{The Eleventh Annual Conference on Large Hadron Collider Physics (LHCP2023)\\
 22-26 May 2023\\
 Belgrade, Serbia\\}

\begin{document}
\maketitle

\section{Introduction and discussion of relevant contributions}
Despite its tremendous success in the description of the physics at collider experiments, 
the Standard Model (SM) is commonly understood to be only an effective theory at currently probed 
energies and precision. 
As the energy range of experiments will not increase much in the near future, 
effects beyond the SM (BSM) are to be observed in the precision domain. 
Potential BSM deviations in the Higgs potential have not yet been investigated at high precision, 
for which di-Higgs production is the key process.


The lack of direct BSM signals in the data suggests that the BSM degrees of freedom are 
well separated from the electroweak (EW) scale, which is a scenario consistently described in 
the framework of bottom-up effective field theories (EFTs) in a model-agnostic way. 
Under the assumption of a decoupling BSM scenario that respects the SM symmetries, the low energy effects are expressed by the linear EFT realisation of Standard Model effective field theory (SMEFT)~\cite{Buchmuller:1985jz,Grzadkowski:2010es,Brivio:2017vri}. 
The SMEFT Lagrangian is described by an expansion in canonical dimension where higher order operators are suppressed by 
higher powers of the scale of new physics $\Lambda$, i.e.
\begin{equation}\label{eq:smeft_expansion}
    \LSMEFT=\LSM+\sum_i\frac{\coeff{i}{}}{\Lambda^2}\mathcal{O}_i + \order{\Lambda^{-4}}\eqcomma
\end{equation}
neglecting baryon- and lepton-number violating operators.
For sufficiently large $\Lambda$ the dominant BSM effects are expected to emerge from dimension-6 operators. 


We constrain our investigations to the gluon fusion channel due to the high luminosity of gluons in proton-proton collisions, 
which leads to a cross section that dominates the other main di-Higgs production channels by a factor of 10~\cite{DiMicco:2019ngk}. 
Moreover, we apply an exact flavour symmetry $\mathcal{G}_\text{flavour}=U(2)_q\times U(2)_u\times U(3)_d$ 
that is compatible with the five-flavour scheme of QCD. 
This choice also reflects the importance of the top-quark for many concrete BSM realisations.


SMEFT predictions are evaluated in a mixed expansion in canonical dimension and loops, 
which we combine with a tree-loop classification of Wilson coefficients of the Warsaw basis 
based on the generic assumption of renormalisability and weak coupling in the BSM sector~\cite{Arzt:1994gp,Buchalla:2022vjp}.
Thus, retaining only CP-even operators, the leading SMEFT contribution originates from
\begin{equation}
\baligned
\Delta\LSMEFT^\text{lead}&=\flambda{\CHbox} \LHbox
+ \flambda{\CHD} \LHD
+ \flambda{\CH} \LH 
\\ &+\flambda{\CtH} \LtH
+\flambda{\CHG} \LHG\eqcomma
\ealigned  \label{eq:LagSMEFT_lead}
\end{equation}
which has been calculated and studied at fixed order NLO QCD~\cite{Heinrich:2022idm} for different truncation options of the cross section. 
The combination with the subset of contributions with additional suppression by a loop factor $(16\pi)^{-1}$ originating from
\begin{equation}
\baligned
\Delta\LSMEFT^\text{sublead}&\supset
\flambda{\CtG} \LtG
+\flambda{\Ctt} \Ltt
\\ &
+\flambda{\CQt} \LQt
+\flambda{\CQto} \LQto
\\ &
+\flambda{\CQQ} \LQQ
+\flambda{\CQQo} \LQQo
\eqcomma  \label{eq:LagSMEFT_chromo4t}
\ealigned
\end{equation}
has been described in Ref.~\cite{Heinrich:2023rsd}.
For reference, we also list the relevant Lagrangian terms of the non-linear EFT realisation (HEFT) contributing to di-Higgs production
\begin{equation}
  \Delta{\cal L}_{\text{HEFT}}=
  -m_t\left(c_{tth}\frac{h}{v}+c_{tthh}\frac{h^2}{v^2}\right)\,\bar{t}\,t -
  c_{hhh} \frac{m_h^2}{2v} h^3+\frac{\alpha_s}{8\pi} \left( c_{ggh} \frac{h}{v}+
  c_{gghh}\frac{h^2}{v^2}  \right)\, G^a_{\mu \nu} G^{a,\mu \nu}\eqcomma
\label{eq:ewchl}
\end{equation}
whose contributions have been thoroughly investigated in Refs.~\cite{Buchalla:2018yce,Heinrich:2020ckp}.

\section{Distributions of leading and subleading contributions}
In this section we present sample diagrams of di-Higgs invariant mass ($m_{hh}$)-distributions generated using the \href{https://powhegbox.mib.infn.it/}{POWHEG-BOX-V2} code \texttt{ggHH\_SMEFT}~\cite{Heinrich:2022idm,Heinrich:2023rsd} and mention sources of theory uncertainties one needs to be aware of.

\begin{table}[htb]
  \begin{center}
   \renewcommand*{\arraystretch}{1.3}
    \begin{footnotesize}
\begin{tabular}{ |c|c|c|c|c|c||c|c|c|c| }
\hline
benchmark  & $c_{hhh}$ & $c_{tth}$ & $ c_{tthh} $ & $c_{ggh}$ & $c_{gghh}$ & $\CHkin$ & $\CH$ & $\CtH$ & $\CHG$ \\
\hline
SM & $1$ & $1$ & $0$ & $0$ & $0$ & $0$ & $0$ & $0$ & $0$ \\
\hline
$1$ & $5.105$ & $1.1$ & $0$ & $0$ & $0$ & $4.95$ & $-6.81$ & $3.28$ & $0$ \\
\hline
\end{tabular}
\end{footnotesize}
\end{center}
  \caption{\label{tab:benchmarks}Definition of benchmark scenarios (with $\CHkin=\CHbox-\CHD/4$). 
  Benchmark point 1 refers to the set in Refs.~\cite{Heinrich:2022idm,Alasfar:2023xpc}, which is an updated version of Ref.~\cite{Capozi:2019xsi}. 
  The parameters were originally derived in HEFT and (naively) translated to SMEFT for $\Lambda=1\;$TeV.} 
\end{table}

\begin{figure}[h]
\includegraphics[width=17pc,page=1]{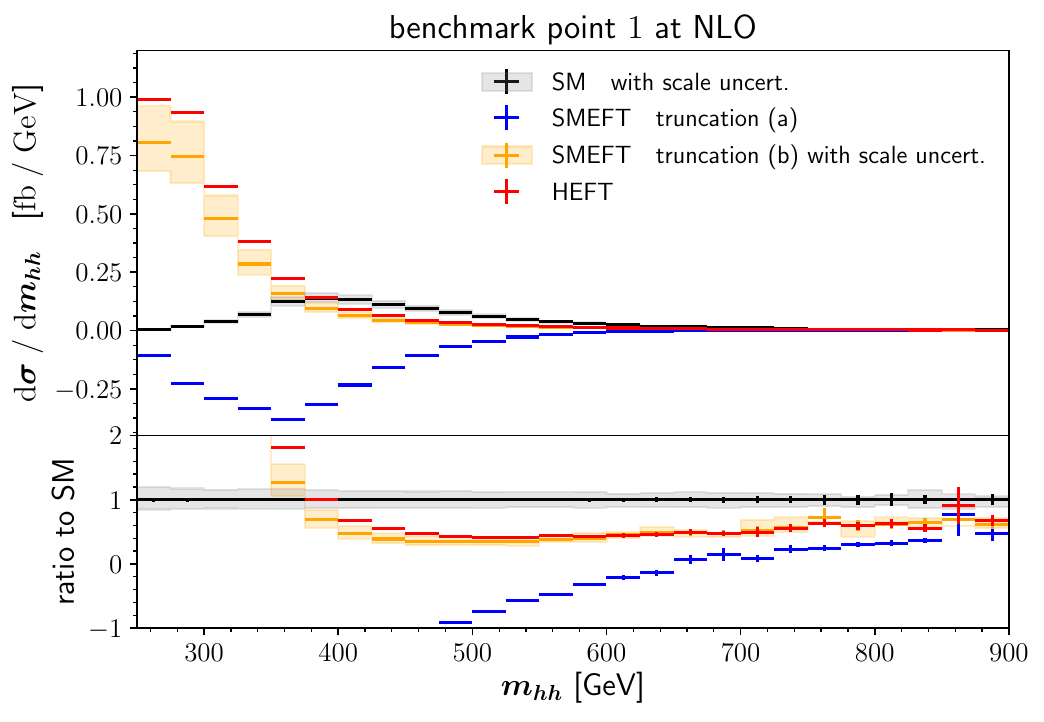}\hspace{1pc}%
\includegraphics[width=17pc,page=1]{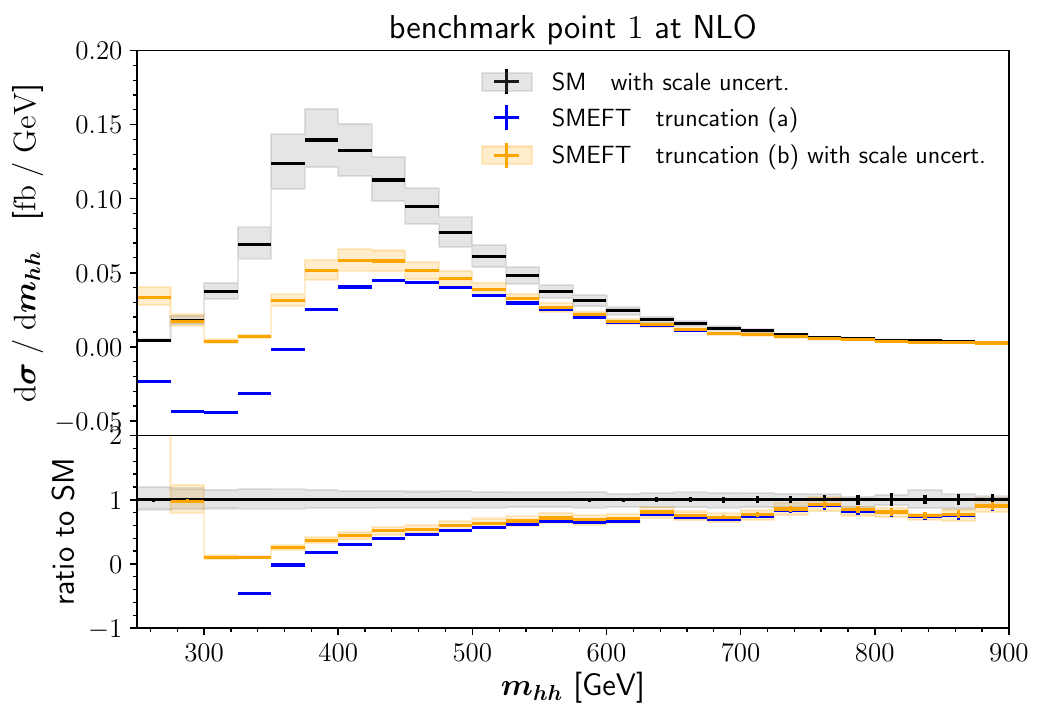}\hspace{1pc}%
   \caption{\label{fig:bpdistributions} Differential cross section distributions for the invariant mass $\mhh$ for benchmark point 1 defined in Table~\ref{tab:benchmarks}. 
   Left: $\Lambda=1$\;TeV, right: $\Lambda=2$\;TeV.}
\end{figure}

In Fig.\ref{fig:bpdistributions}, the distributions for benchmark 1 defined in Table~\ref{tab:benchmarks} are displayed for $\Lambda=1,\,2\;$TeV 
and for the linear (a) and linear+quadratic (b) truncation options of the cross section. The SM and HEFT distributions are shown for comparison. 
The negative cross section for truncation option (a) demonstrates that the naive translation from a valid HEFT point to SMEFT leads 
to parameter points incompatible with a truncation at canonical dimension-6, highlighting the importance to study HEFT and SMEFT separately. 
Moreover, approaching the SM configuration with increasing value of $\Lambda$ shows a convergence of the difference between the truncation options, 
which reflects the expected behaviour of being a qualitative proxy for the estimation of the uncertainty related to the SMEFT truncation. 

\begin{figure}[h!]
\includegraphics[width=16pc,page=1]{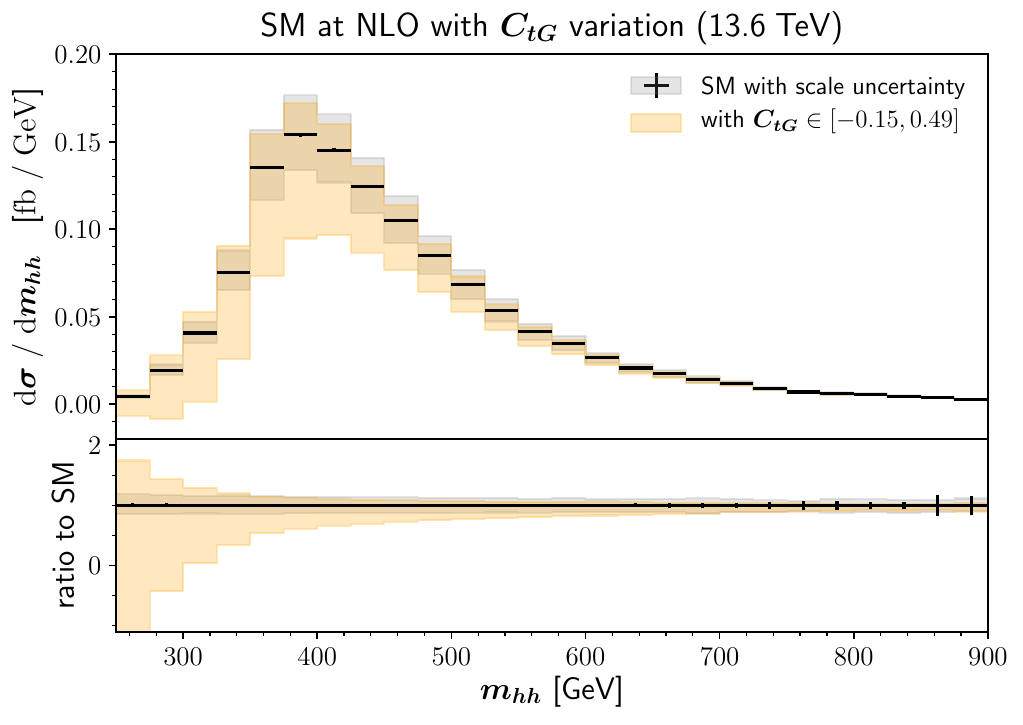}\hspace{3pc}%
\includegraphics[width=16pc,page=1]{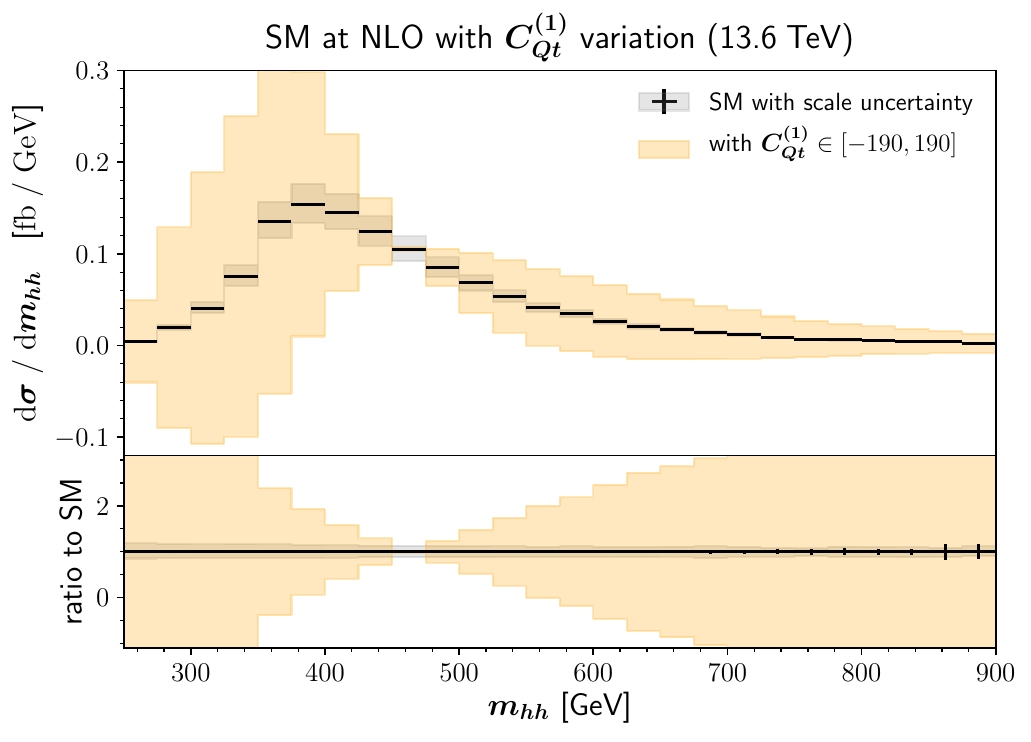}\hspace{3pc}%
   \caption{\label{fig:subleaddistributions} Differential cross section for 
   a variation of $\CtG$ (left) and $\CQt$ (right) w.r.t. the SM $\mhh$-distribution. 
   The ranges are taken from Ref.~\protect\cite{Ethier:2021bye} based on a marginalised ${\cal O}(\Lambda^{-2})$ fit.}
\end{figure}

In Fig~\ref{fig:subleaddistributions} we present the variation of $\CtG$ and $\CQt$ using conservative bounds from marginalised fits of Ref.~\cite{Ethier:2021bye}. 
Variations of the other Wilson coefficients in Eq.~\eqref{eq:LagSMEFT_chromo4t} can be found in Ref.~\cite{Heinrich:2023rsd} 
which also discusses the effect of different $\gamma_5$ schemes following the derivation of Ref.~\cite{DiNoi:2023ygk}. 
The observed deviations from the SM curve together with the study of Ref.~\cite{Alasfar:2022zyr} demonstrate the potential 
for improved limits for $\CQt$ and $\CQto$ in global fits when indirect effects in single and di-Higgs production are included.

%
In the following, we briefly list the relevant sources of theory uncertainties which are described in more detail in Ref.~\cite{Alasfar:2023xpc}:
\begin{description}[leftmargin=!,labelwidth=0.5cm]
    \itemsep0em 
    \item[SMEFT truncation:] There is no quantitative prescription available yet~\cite{Brivio:2022pyi}. It is possible to get a qualitative picture comparing different truncation options as proxy for each EFT point~\cite{Heinrich:2022idm}.

    \item[Scale uncertainty:] The scale uncertainty is assessed by a variation of renormalisation and factorisation scales around the central choice $\mu_0=\mhh/2$.

    \item[PDF+$\alpha_s$ uncertainty:] Estimated to be about $\pm3\%$ for $\sqrt{s}=13$ and $14\;$TeV, which appears to be
    robust for $\chhh$-variations~\cite{HHTwiki}.

    \item[$m_t$ renormalisation scheme:] Comparison between on-shell and $\MSbar$ for different scales results in $_{-18\%}^{+4\%}$ for the SM at $\sqrt{s}=13\;$TeV, 
    a dependency on $\chhh$ and bin width has been observed~\cite{Baglio:2020wgt,Bagnaschi:2023rbx}.

    \item[EW corrections:] Have been calculated for SM~\cite{Bi:2023bnq,Davies:2023npk}, but are not translatable to SMEFT.
    
    \item[NLO QCD virtual corrections:] 
    The two-loop virtual corrections are encoded in numerical grids based on the distributions of events in the SM. For SMEFT scenarios where the low-$m_{hh}$ region or the tail of the $m_{hh}$-distribution is much more populated than in the SM, large statistical uncertainties can arise due to an insufficient number of grid points in the region.
    
\end{description}

\section{Conclusions}
We presented results of state-of-the-art predictions for di-Higgs production in SMEFT, 
pointed out 
different options for the truncation of the EFT expansion and the inclusion of subleading operators 
and briefly mentioned the remaining theory uncertainties. 
An important outstanding task is the inclusion renormalisation group evolution effects, as they are expected to be sizable following recent results for other processes~\cite{Battaglia:2021nys,Aoude:2022aro,DiNoi:2023onw}. 

\section*{Acknowledgements}

This research was supported by the Deutsche Forschungsgemeinschaft (DFG, German Research Foundation) under grant 396021762 - TRR 257.

\bibliographystyle{JHEP}

\bibliography{HH_preciseSMEFT_LHCP_Main}


\end{document}